\documentclass{emulateapj}
\received{}
\revised{}
\accepted{}
\usepackage{amsmath}
\usepackage{amssymb}
\usepackage{graphics}
\usepackage{graphicx}
\usepackage{natbib}
\usepackage{gensymb}
\usepackage{capt-of}
\usepackage{morefloats}
\usepackage{xcolor}
\usepackage{floatrow}
\usepackage{appendix}
\usepackage{cleveref}
\shorttitle{Doppler velocities in transition region active regions}
\shortauthors{Ghosh et. al.}

\begin{document}

\title{On Doppler shift and its Center-to-Limb Variation in Active Regions in the Transition Region}
\author{Avyarthana Ghosh\altaffilmark{1,2}, James A. Klimchuk\altaffilmark{3}, Durgesh Tripathi\altaffilmark{1}}
\affil{$^1$Inter-University Centre for Astronomy and Astrophysics, Post Bag - 4, Ganeshkhind, Pune 411007, India}
\affil{$^2$Center of Excellence in Space Sciences India, Indian Institute of Science Education and Research, Kolkata, West Bengal 741246, India}
\affil{$^3$NASA Goddard Space Flight Center, Code 671, Greenbelt, MD 20771, USA}
\date{}
\begin{abstract}

A comprehensive understanding of the structure of Doppler motions in transition region including the center-to-limb variation and its relationship with the magnetic field structure is vital for the understanding of mass and energy transfer in the solar atmosphere. In this paper, we have performed such a study in an active region using the Si IV 1394~{\AA} emission line recorded by the Interface Region Imaging Spectrograph (IRIS) and the  line-of-sight photospheric magnetic field obtained by the Helioseismic and Magnetic Imager (HMI) on-board the Solar Dynamics Observatory (SDO). The active region has two opposite polarity strong field regions separated by a weak field corridor, which widened as the active region evolved. On average the strong field regions (corridor) show(s) redshifts of 5{--}10 (3{--}9)~km~s$^{-1}$ (depending on the date of observation). There is, however, a narrow lane in the middle of the corridor with near-zero Doppler shifts at all disk positions, suggesting that any flows there are very slow. The Doppler velocity distributions in the corridor seem to have two components---a low velocity component centered near 0 km/s and a high velocity component centered near 10~km~s$^{-1}$. The high velocity component is similar to the velocity distributions in the strong field regions, which have just one component. Both exhibit a small center-to limb variation and seem to come from the same population of flows. To explain these results, we suggest that the emission from the lower transition region comes primarily from warm type II spicules, and we introduce the idea of a `chromospheric wall'---associated with classical cold spicules---to account for a diminished center-to-limb variation.

\end{abstract}

\keywords{Sun: activity -- Sun: photosphere -- Sun: transition region -- Sun: magnetic fields -- Sun: sunspots}

\section{Introduction}

Active regions are locations of concentrated complex magnetic structures on the Sun's surface. They demonstrate pronounced coronal heating \citep[see][and references therein]{Rea_2014} as well as flares and coronal mass ejections \citep{TriBC_2004, WebH_2012, Ben_2017}. Therefore, it is mandatory to comprehend the local as well as global Doppler motions in active regions and compare those with the structure of the magnetic field. This in turn will help us understand the transfer of mass and energy in the solar atmosphere.

Some of the earliest observations of Doppler velocities in regions of different magnetic field strengths and configurations have been done with the Orbiting Solar Observatory (OSO-8) \citep{Bru_1977}, the Naval Research Laboratory (NRL) normal incidence spectrograph on Skylab (S082-B), the NRL High Resolution Telescope and Spectrograph \citep[HRTS;][]{BarB_1975} and the Ultraviolet Spectrometer and Polarimeter \citep[UVSP;][]{WooTB_1980} on-board the Solar Maximum Mission \citep[SMM;][]{Sim_1981}. Using these early days observations, downflows as large as 10{--}20~km~s$^{-1}$ in ultraviolet (UV) spectral lines over large-scale plage regions have been reported \citep[see e.g.,][]{LemSA_1978, GebHT_1980, BruBD_1980, Lit_1980, Bru_1981, AthGH_1982, Der_1982, RotOK_1982, AthGH_1983b, Bre_1993, AchBK_1995}. In some cases, redshifts of magnitudes $\sim$80{--}100~km~s$^{-1}$ in transition region spectral lines over small isolated patches of active regions have been reported in the observations recorded using the NRL HRTS \citep{NicBB_1982, DerBB_1984}. However, such high downflows are relatively uncommon and have been attributed to dynamic/explosive events.

\cite{Kli_1987} performed one of the first comprehensive studies of Doppler velocities in the transition region using UVSP data and compared them with the structure of the photospheric magnetic field obtained from the Kitt Peak National Solar Observatory (NSO). This study was conducted for 25 active regions with a variety of locations across the solar disk. It was found that the strong field regions were relatively redshifted by 5{--}10~km~s$^{-1}$ irrespective of disk position and were approximately steady over a spatial scale of $\geq$~3{\arcsec}. The weak field corridors separating strong field regions of opposite polarity were relatively blueshifted. However, no absolute velocity reference was available, and the red and blueshifts are relative to the 4x4 arcmin$^{2}$ raster averages. As discussed later, \cite{Kli_1986a, Kli_1987} argued that the weak field corridors likely have an absolute velocity near 0~km~s$^{-1}$, independent of disk position.

In more recent times, similar observations continued employing other high resolution facilities like the Solar Ultraviolet Measurements of Emitted Radiation \citep[SUMER;][]{WilCM_1995} and the Coronal Diagnostic Spectrometer \citep[CDS;][]{HarSC_1995} on-board the SOlar and Heliospheric Observatory \citep[SOHO;][]{DomFP_1995}, the EUV Imaging Spectrometer \citep[EIS;][]{CulHJ_2007} on-board Hinode \citep{KosMS_2007} and the Interface Region Imaging Spectrograph \citep[IRIS;][]{DePTL_2014}. \cite{TerBD_1999} (also see the references therein) used SUMER observations to ascertain that the downflows in active regions increase from $\sim$0~km~s$^{-1}$ at $\log\,T[K]=$ 4.3 to about 15~km~s$^{-1}$ at $\log\,T[K]=$ 5.0 that changes to $\sim$8~km~s$^{-1}$ blueshifts in the \ion{Ne}{8} line at $\log\,T[K]=$ 5.8. The observations from the EIS have further augmented our understanding of plasma flows in various structures in active regions such as loops and moss \citep[see for $e.g.,$][and the references and citations therein]{Del_2008, BroW_2009, TriMD_2009, DadTS_2011, TriMK_2012, GupTM_2015, GhoTG_2017}. However, we note that for most EIS observations, the lower transition region lines $viz.,$ \ion{O}{4}, \ion{O}{5} and \ion{Mg}{5} are very weak \citep{YouZM_2007}. Hence we have a few conclusive studies of Doppler shifts in the lower transition region.

In order to compare the flow structure with magnetic field structures, Center-to-Limb Variation (CLV) studies of flows have been carried out \citep[$e.g.,$][]{RotHJ_1990, HasRO_1991a}. \cite{FelDC_1982} used observations of two active regions as these traversed across the solar disk by the NRL normal incidence spectrograph on Skylab (S082-B) to show that the downflows typically range between 4{--}17~km~s$^{-1}$ (for spectral lines formed between $\log\,T[K]=$ 4.7{--}5.0). However, there is a slight tendency of decrease near the limb as compared to that on the disk. However, the plasma has significantly lower magnitudes of Line-Of-Sight (LOS) velocities for spectral lines formed at $\log\,T[K]>$ 5.0. This study includes wavelength calibration with respect to neutral and singly ionized spectral lines formed in the chromosphere (\ion{C}{1}, \ion{O}{1} and \ion{Si}{2}). A series of studies by \cite{RouS_1982, RotHJ_1990, HasRO_1991a} show a decrease of downflows from a few km~s$^{-1}$ at the center to about 0~km~s$^{-1}$ towards the limb in quiet Sun region. It is imperative to mention that for wavelength calibration, \cite{RouS_1982} used cooler choromospheric lines (\ion{Si}{1} and \ion{Fe}{2}) as well as comparison of on-disk and near-limb spectra and raster averages at different locations on the solar disk. On contrary,  \cite{RotHJ_1990, HasRO_1991a} used on-board sources to derive the reference.

To perform a comprehensive study of Doppler shifts and its CLV, we use the excellent observations over an active region recorded by IRIS in a \ion{Si}{4} line as the region crossed through the central meridian. We note that this is the best dataset available for such an analysis. The magnitudes and directions of the plasma flows are compared with photospheric magnetic field structure observed by the Helioseismic Magnetic Imager \citep[HMI;][]{SchBN_2012,SchSB_2012} on-board the Solar Dynamic Observatory \citep[SDO;][]{PesTC_2012}. The presence of several neutral and singly ionized spectral lines in IRIS spectra gives us the chance to perform absolute wavelength calibration, which was missing in the studies performed earlier, such as the one by \cite{Kli_1987}. These observations are in tandem with observations taken by the Atmospheric Imaging Assembly \citep[AIA;][]{LemTA_2012}, also on-board SDO. The rest of the paper is structured as follows. In \S\ref{obs}, we provide a detailed description of the data used, a brief description of the instruments used and discuss the processing techniques. In \S\ref{analy}, we present the method of analysis and results, followed by a summary in \S\ref{sum}. Finally, we present a possible interpretation of the observations in the discussion in \S\ref{disc}. 

\section{Observations} \label{obs}

\begin{figure*}[!hbtp]
\centering
\includegraphics[trim=0.cm 0.cm 0.cm 0.cm, width= 0.8\textwidth]{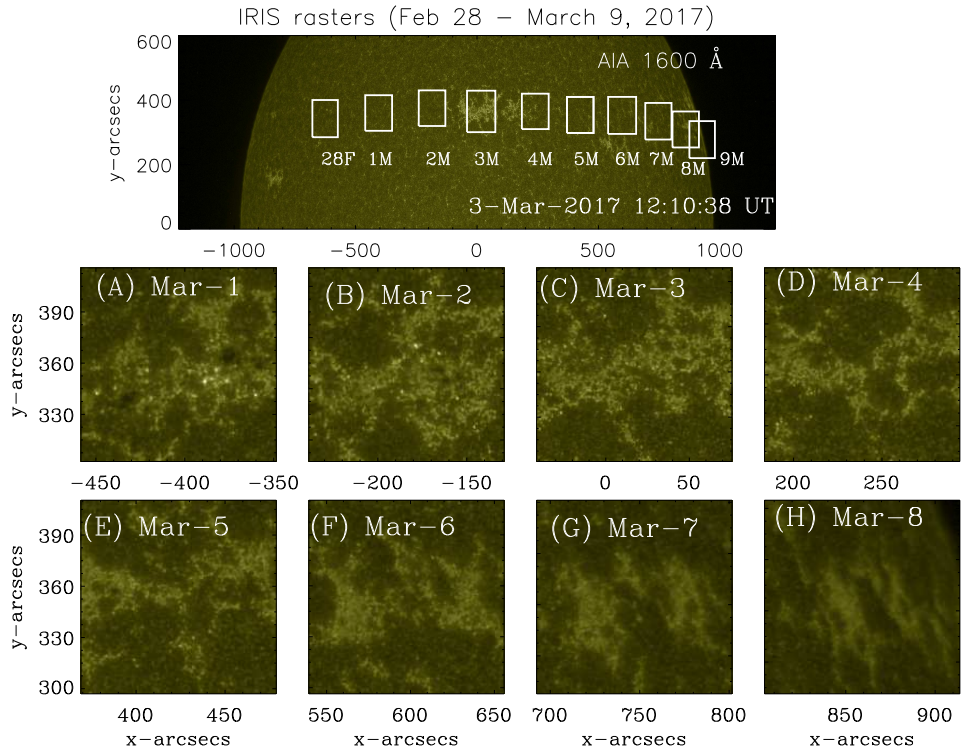}
\caption{Top panel: Location of the IRIS raster field of view (shown in white boxes) from 28th Feb until March 9th of active region \textsl{AR 12641} on a partial disk image of the Sun taken in the 1600~{\AA} channel of AIA on March 3, 2017. The letter `F' stands for February and `M' for March. Middle and bottom panels: Zoomed-in views of the IRIS raster FOVs on between March 1{--}8 as seen in AIA 1600~{\AA} channel.}\label{track}
\end{figure*}

Active region \textsl{AR~12641} was located in the northern hemisphere and tracked by the IRIS across the solar disk between February 28 and March 9, 2017. For this purpose a 320-step dense raster was used that had an exposure time of 4 seconds. The instrument has a very high spatial resolution of 0$\arcsec$.4, which corresponds to less than 300~kms near the disk-center. The top panel of Fig.~\ref{track} displays the locations and Fields-Of-View (FOV) on different days (as labelled), over-plotted on a partial disk image recorded by the AIA in 1600~{\AA} channel on March~3 when the active region was located at the central meridian. We further display zoomed-in views of the FOVs from March 1{--}8, 2017 as observed by AIA 1600~{\AA} in the middle and bottom rows (marked A-H) of Fig.~\ref{track}. Here, we note that on March 1, there are two very small sunspots of opposite polarity. These disperse on the following days, but a large scattered plage region persists on all successive days. The plage region changes its shape as it traverses across the disk. The first and second columns of Table~\ref{fe2_t} provide the date and $\mu$-value of each observation (where $\mu$ is defined as the cosine of the heliographic longitude) of the center of the FOV.

For this work, we have used Level-2 IRIS raster data which is corrected for all instrumental effects such as flat-fielding, dark currents, offsets and thermal orbital variations so as to make it suitable for scientific analysis\footnote{A User’s Guide To IRIS Data Retrieval, Reduction $\&$ Analysis, S.W. McIntosh, February 2014}. The IRIS data are analysed using Gaussian fitting routines provided in \textsl{Solarsoft}\footnote{Using EIS Gaussian fitting routines for IRIS data, P. Young, April 2014} \citep{FreH_1998}.

For comparing the structure of the Doppler shifts with that of magnetic field in the photosphere, we have used the photospheric LOS magnetograms obtained by the HMI on-board SDO. In order to infer the overlying coronal structures, we have also used the 171~{\AA} images recorded by AIA.

\section{Data Analysis and Results} \label{analy}

The IRIS FUV2 spectra records a pair of \ion{Si}{4} lines at 1393.78~{\AA} and 1402.77~{\AA} ($\log\,T[K] =$  4.9) that can be used to measure the Doppler velocities in the lower transition region. In optically thin plasma conditions, \ion{Si}{4} 1393.78~{\AA} is a factor of 2 stronger than 1402.77~{\AA}, as derived using CHIANTI \citep{DerML_1996, LanYD_2013}. Since, our aim is to measure the Doppler shift, it is preferable to use the stronger line at 1393.78~{\AA}. 

\begin{center}
\captionof{table}{Table giving the dates and $\mu$ values for the IRIS observations, followed by the central wavelengths of the \ion{Fe}{2} line as determined from single Gaussian fits of the profiles without and with ICSF correction (columns 3 and 4), respectively.}\label{fe2_t}
\label{details}
\vspace{1cm}
\begin{tabular}{| c | c | c | c | }  \hline   
Date &$\mu$ &Central wavelength  &Central  \\
2017 &   &without    &wavelength \\ 
 & &correction ({\AA}) & with ICSF ({\AA}) \\ \hline
28 Feb &0.66 &-&- \\
Mar 1  &0.82 	&1392.818	&1392.823 \\
Mar 2  &0.90 	&1392.810	&1392.817 \\
Mar 3  &0.92 	&1392.818	&1392.820 \\
Mar 4  &0.89 	&1392.820	&1392.820 \\
Mar 5  &0.81 	&1392.810	&1392.816 \\
Mar 6  &0.68 	&1392.826	&1392.830 \\
Mar 7  &0.51 	&1392.815	&1392.812  \\
8 Mar &0.27 &1392.826 &1392.828 \\
9 Mar &0.00 &-&- \\ \hline
\end{tabular}
\end{center}


\subsection{Wavelength Calibration}\label{fe2}

\begin{figure*}[!hbtp]
\centering
\includegraphics[trim=0.cm 1.cm 0.cm 0.cm,width= 1.0\textwidth]{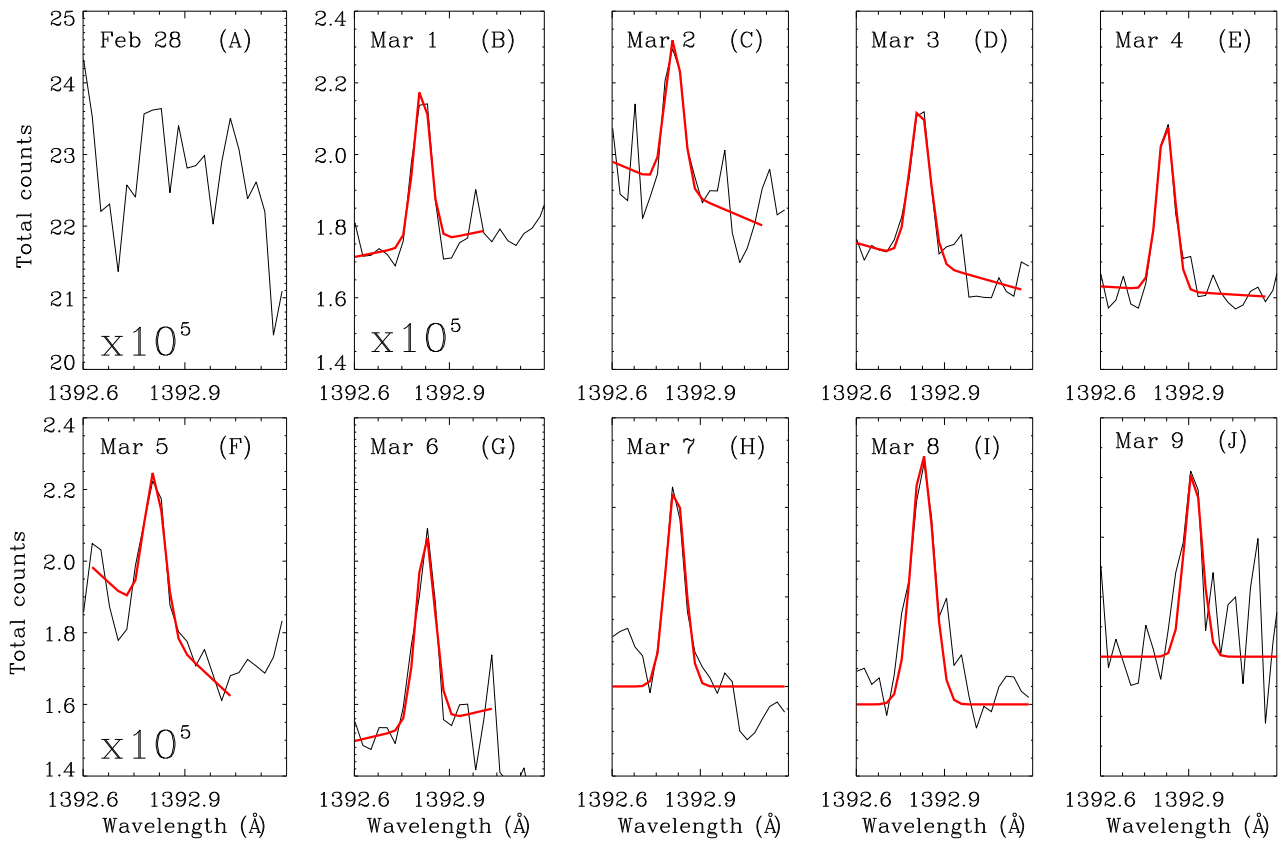}
\caption{\ion{Fe}{2}~1392~{\AA} spectra averaged over the entire FOV taken between February 28, 2017{--}March 9, 2017. The original profiles are shown in black and fitted Gaussians are shown in red.}\label{calib1}
\end{figure*}

In the absence of an on-board lamp, an absolute wavelength calibration can be derived using cooler neutral or singly ionised spectral lines formed in the photosphere or chromosphere, which are considered to be at rest \citep{HasRO_1991b}. Such lines, available in the IRIS spectral window are, \ion{S}{1}, \ion{O}{1} and \ion{Fe}{2}. Generally, in order to minimise errors associated with uncertainties due to non-linear dispersion, it is better to use a spectral line for calibration which is formed at a wavelength closer to the spectral line being used for velocity measurements, which is \ion{Si}{4} 1394~{\AA}, in the present case. We, therefore, chose \ion{Fe}{2}~1392.8~{\AA} for our initial calibration. Fig.~\ref{calib1} displays the spectra of \ion{Fe}{2}~1392.8~{\AA} line obtained over the entire IRIS raster FOV for all dates between February 28 to March 9, 2017. As can be seen, the \ion{Fe}{2} line is very well identified on all days except on February 28 (panel A of Fig.~\ref{calib1}). Moreover, on 9 March (J), the active region was almost on the west limb. Hence, we exclude this date for further analysis. 

\begin{figure*}[!hbtp]
\centering
\includegraphics[trim=0.cm 1.cm 0.cm 0.cm, width= 1.0\textwidth]{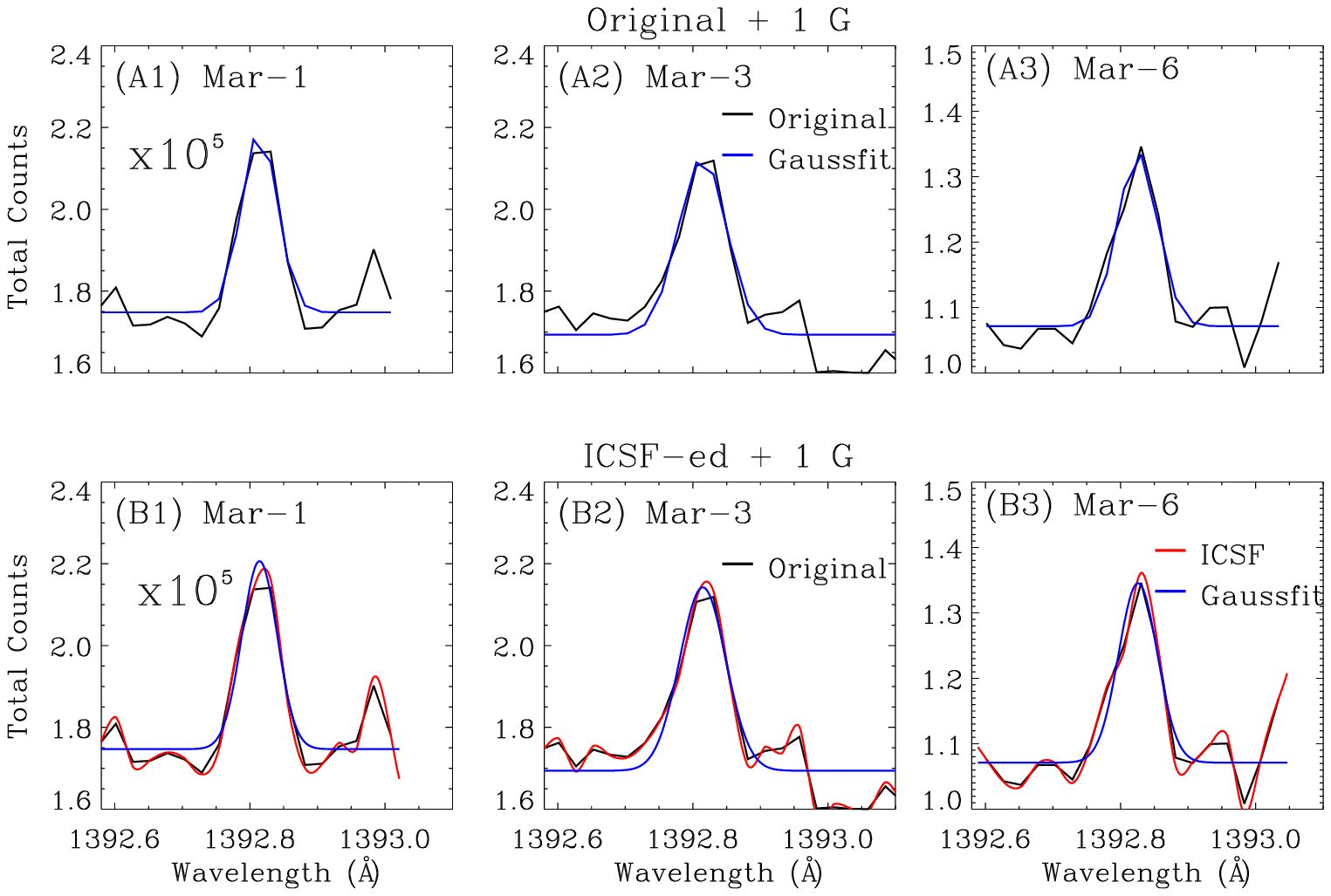}
\caption{Top Row: Original spectra (in black) of \ion{Fe}{2} 1392~{\AA} averaged over the entire FOV over-plotted a single Gaussian fit (in blue). Bottom Row: The original spectra (black) over-plotted with spectra obtained via ICSF (Intensity Conserving Spectral Fitting) in red and fitted with single Gaussian over obtained spectra in blue. The respective dates are mentioned at the top of each panel.}\label{calib2}
\end{figure*}

The black solid lines in the different panels in Fig.~\ref{calib1} are the original \ion{Fe}{2} line profiles, whereas the over-plotted red lines are the corresponding Gaussian fits. It can be seen that the line is not broad compared to the spectral bin size, i.e., there are a limited number of measurements across the profile. We therefore take an additional step in order to measure the line center position as accurately as possible. We apply a procedure called Intensity Conserving Spectral Fitting \citep[ICSF;][]{KliPT_2016}, which accounts for the finite size of the spectral bins. A spectrometer measures the average intensity over a bin, and this intensity is typically assigned to the wavelength position at bin center. This is incorrect, however, when the true line profile is not a straight line within the bin, as it surely is not. ICSF approximates the detailed shape of the true profile using a spline fit, then determines the appropriate intensity at bin center, which is slightly different from the average over the bin. We have applied the ICSF correction to our data and then proceeded to perform a second set of Gaussian fits. The top panels in Fig.~\ref{calib2}, labelled (A), show the original \ion{Fe}{2} line profiles averaged over the entire FOV (in black) on March 1, 3 and 6, where March 3rd data was located at the disk center. The blue curves are single Gaussian fits. Panels (B) on the bottom show the original spectra in black and the ICSF-corrected line profiles in red. The blue curves show the corresponding Gaussian fits to the ICSF-corrected line profiles.

Table~\ref{fe2_t} lists the difference in the obtained central wavelength of the fitted Gaussian before and after the application of ICSF procedure. The maximum difference is on March 2, 2017 which is 7~m{\AA}, translating to $\sim$1.5~km~s$^{-1}$, that is equivalent to uncertainty in the velocities obtained with IRIS \citep{DePTL_2014}, hence significant. However, the laboratory rest wavelength used for absolute wavelength calibration is 1392.817~{\AA} \citep{SanB_1986}, which is 1~m{\AA} larger than that derived using off-limb observations recorded with HRTS \citep[see e.g.,][]{Her_1962, KauW_1966, Per_1971, BroTT_1974}. This difference translates to an uncertainty of $\sim$0.2~km~s$^{-1}$ in velocity measurements. We further note that there is an uncertainty of $\sim$3~m{\AA} (0.66~km~s$^{-1}$) in the HRTS measurements. 

We note in Fig.~\ref{calib2} that the line profile has enhanced emission on the blue side, near 1392.78~{\AA}. It is present on March 1 and 6, when the active region is away from disk center, and even more pronounced on March 8 and 9, when the region is close to the limb. (There is a separate and possibly unrelated enhancement farther out in the blue wing, near 1392.75~{\AA}, in the disk center observation of March 3). We suggest that the asymmetry in the shape of the \ion{Fe}{2} profile may be due to opacity effects. A longer path length near the limb would be associated with a greater optical thickness, thereby producing a stronger self-reversal. Previous literature suggests that cooler lines $viz.$ \ion{C}{2}, \ion{C}{3}, \ion{O}{1}, \ion{Si}{2}, \ion{Si}{3}, \ion{S}{1} $etc.$ are more prone to opacity effects \citep{FelDP_1976, DoyW_1980}. This could also be true for \ion{Fe}{2} line which is formed at similar temperatures. Indeed, \cite{PolDD_2016} used IRIS spectra over a number of features with varied magnetic field topologies to find very different line shapes and widths, some of which could be fitted with non-Maxwellian $\kappa$-distribution instead of a Gaussian fitting \citep{DudPD_2017}.

\begin{figure*}[!t]
\centering
\includegraphics[width= 1.0\textwidth]{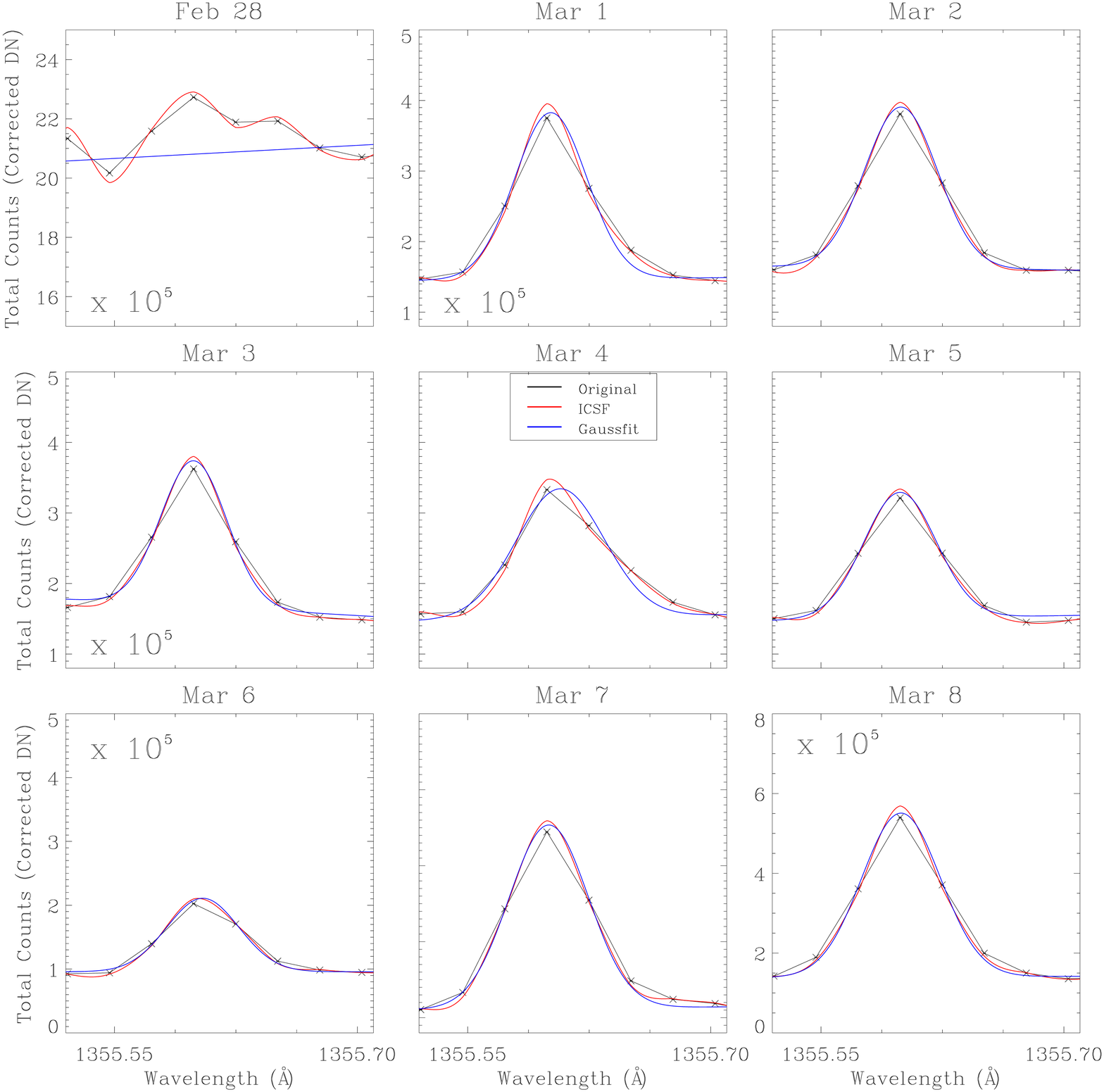}
\caption{Original spectra of \ion{O}{1} 1356~{\AA} averaged over the entire raster (black) over-plotted with spectra obtained through ICSF (red) and the fitted Gaussian profiles (blue). The maximum error on these profiles for all these days does not exceed 0.40\%, assuming only photon shot noise.}\label{calib3}
\end{figure*}

Because the shape of the \ion{Fe}{2} profile and its CLV are not understood, we decided to consider another line for wavelength calibration. The other choices are \ion{O}{1} 1355.6~{\AA} and \ion{S}{1} 1401.52~{\AA}. \ion{S}{1} is closer in wavelength to \ion{Si}{4}, but it is very faint in all our observations, so we settled on \ion{O}{1}, which is an optically thin mid-chromospheric line \citep{LinC_2015}. Bart De Pontieu (private communication) has informed us that the IRIS dispersion is known well enough that the wavelength separation should not be a problem. Fig.~\ref{calib3} shows the line profiles of \ion{O}{1} for all the days between February 28 to March 8. The lower limit of error on these profiles (total counts) for all these days does not exceed 0.40\%, assuming only photon shot noise. With the exception of February 28, the profiles look ordinary, with no blue wing enhancement, most likely due to the fact that it is an optically thin line \citep{LinC_2015}. Therefore, we consider it as a well suited line for our wavelength calibration. The laboratory rest wavelength for \ion{O}{1} is 1355.598~{\AA} \citep{SanB_1986}. The second column of Table~\ref{tab_dv} lists the central wavelengths obtained from Gaussian fitting of the \ion{O}{1} line profiles after ICSF correction. Note that the central wavelengths obtained for March 4 and 6 are comparatively larger than that for other dates. Such differences could be attributed to the original line profiles (refer to the respective panels in Figure~\ref{calib3}). We further note that the wavelength calibration is accurate to 0.67~km~s$^{-1}$. 

\subsection{Co-alignment and distinction of weak and strong field regions}\label{mag}

Since we are dealing with more than one instrument in this study i.e., IRIS and HMI, a proper co-alignment between the observations taken from different vantage points is mandatory. In this work, the HMI LOS magnetogram at a given time is co-aligned with the AIA 1600~{\AA} image that is further co-aligned with IRIS \ion{Si}{4} raster images. The scheme of co-alignment was performed separately for each data set taken from March 1{--}8, 2017.

\begin{figure}[t!]
\centering
\includegraphics[trim=1.cm 0.cm 0.cm 0.cm,  width= 1.1\linewidth]{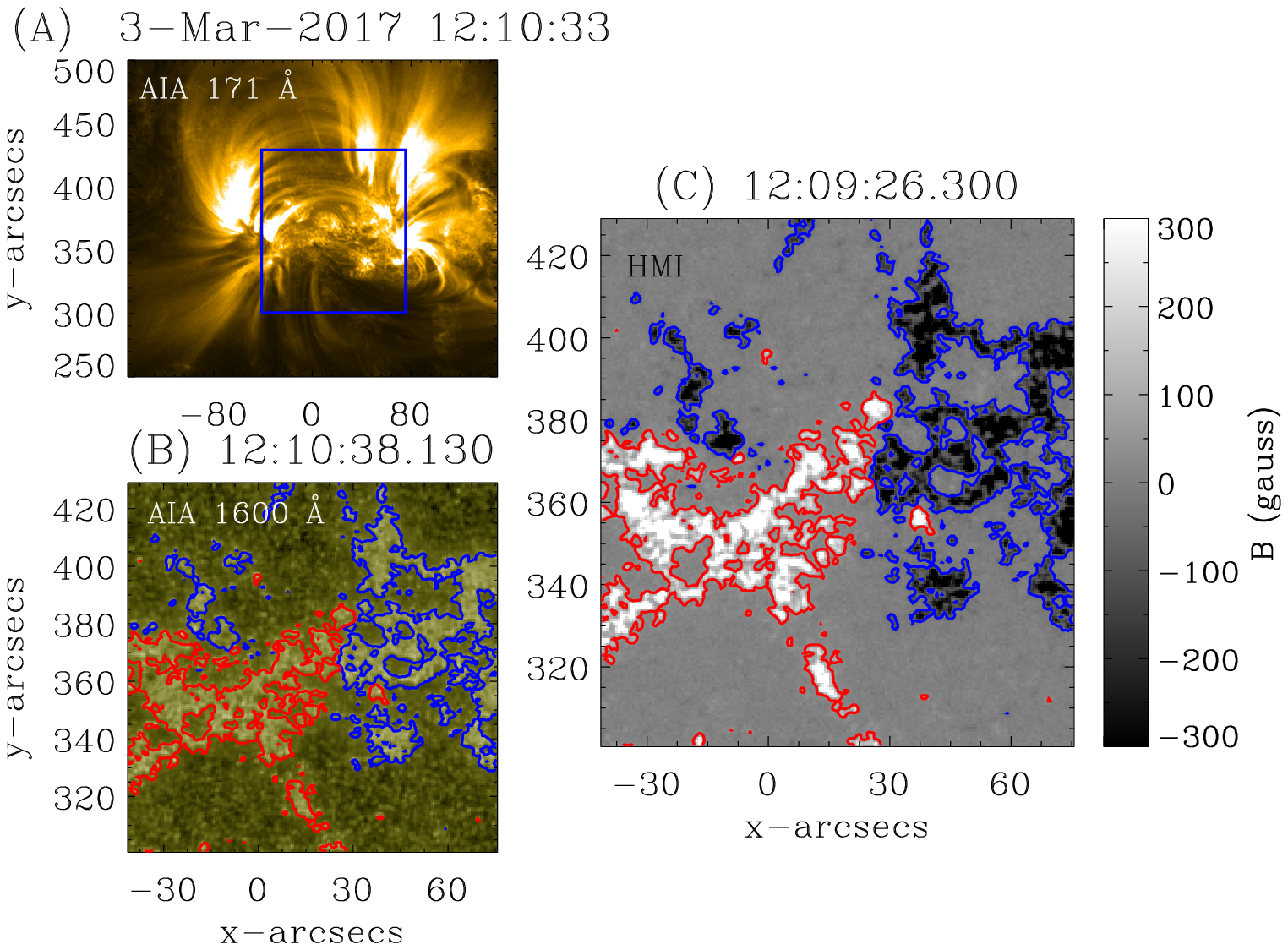}
\caption{Panel (A): AIA 171~{\AA} image taken on March 3, 2017 over-plotted with IRIS field of view (blue box). Fan loops can be seen emanating as well. Panel (B): The AIA 1600~{\AA} intensity map. Panel (C): HMI LOS magnetic field maps. In both panels (B) \& (C) (taken at nearly same time), the red and blue contours show magnetic fields of levels of 50~G and -50~G, respectively.}\label{location}
\end{figure}

\cite{Kli_1987} distinguished strong and weak field regions based on a LOS field strength of 100~G, i.e., strong field regions are enclosed by $\pm$100~G contours. In this study, we use a field strength of 50~G. The choice is not critical since the spatial gradient of the LOS field is quite steep at these values. Contours at 50 and 100~G are closely spaced. Furthermore, the uncertainty in the measurement of the photospheric magnetic field is about 17~G for 50 seconds observation\footnote{Instrument Performance Document, Helioseismic and Magnetic Imager for Solar Dynamic Observatory, CDR Version 16 November 2004} for HMI. The strip that separates strong field regions of positive and negative polarity in an active region is referred to as the weak field corridor, following \cite{Kli_1987}.

Panel (A) of Fig.~\ref{location} displays the active region as observed in AIA 171~{\AA} with the superimposed blue box indicating the region that was rastered with IRIS on March 3, which was closest to the disk center. Panel (B) shows the region-of-interest (ROI) with AIA 1600~{\AA} channel whereas panel (C) shows the photospheric magnetic field obtained with HMI. The overlaid red and blue contours represent $\pm$50~G levels of LOS magnetic field with red being positive and blue being negative polarity fields after co-alignment. The contours with levels $\pm$50~G cover almost all the strong polarity regions in the active regions. These contours in panel (B) over AIA~1600{\AA} image shows excellent overlap with the plage regions, which are essentially the footpoints of coronal loops like those seen in the 171~{\AA} image (panel A).

\subsection{Doppler velocities in strong field regions}\label{dv_str}

\begin{figure*}
\centering
\includegraphics[trim=0.cm 4.cm 0.cm 1.cm, width= 1.0\textwidth]{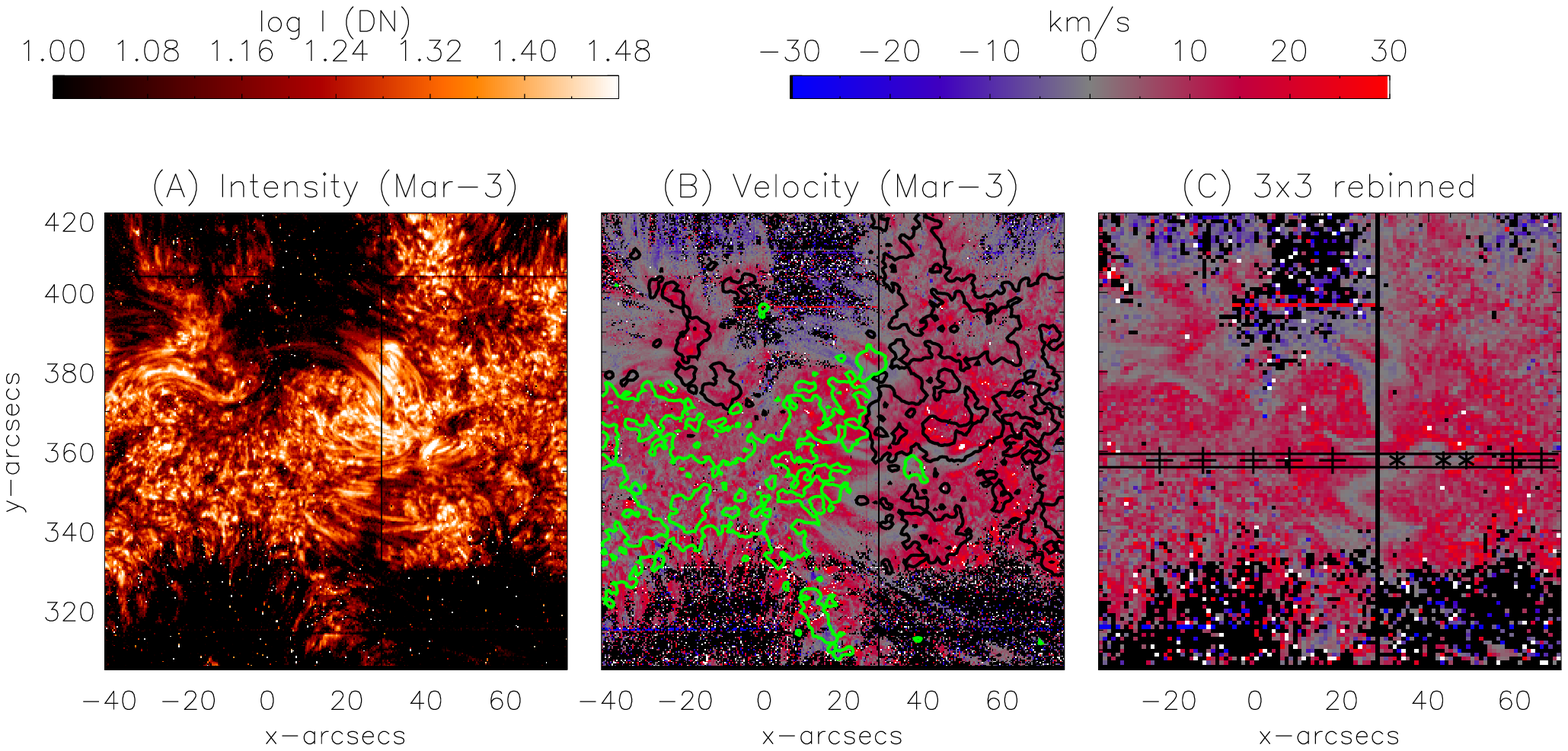}
\caption{\ion{Si}{4}~1394~{\AA} intensity (panel A) and velocity map (panel B) on March 3. The over-plotted green (black) contours in panel (B) correspond to 50 (-50)~G obtained from HMI. Panel (C) shows the velocity map binned over 3 $\times$ 3 pixels. The black pixels represent the missing data as well as upflows greater than 30~km~s$^{-1}$. The white pixels show the downflows exceeding 30~km~s$^{-1}$. Over-plotted is a horizontal stripe containing `+' for strong field regions and `*' in the corridor region.}\label{march3}
\end{figure*}

Fig.~\ref{march3} displays \ion{Si}{4}~1394~{\AA} intensity (panel A) and velocity maps displayed within the range of $\pm$30~km~s$^{-1}$ (panels B \& C). In panel B, we plot the $\pm$50~G contours of the photospheric magnetic field. The velocity maps show predominant redshifts. A clear exception is a `skiing-track' like feature with very small velocities. This track corresponds to the weak field corridor separating the two dominant magnetic polarities. Velocities are also diminished in the weak field areas surrounding the active region. Note that the black pixels in the velocity maps represent bad pixels (no measurement) as well as saturated pixels (blueshifts exceeding 30~km~s$^{-1}$). Further, the white pixels denote redshifts exceeding 30~km~s$^{-1}$. In Panel (C), the same velocity map is plotted with 3 $\times$ 3 pixels binning.

\begin{figure}[t!]
\centering
\includegraphics[trim=1.cm 0.cm 0.cm 0.cm,  width= 0.95\linewidth]{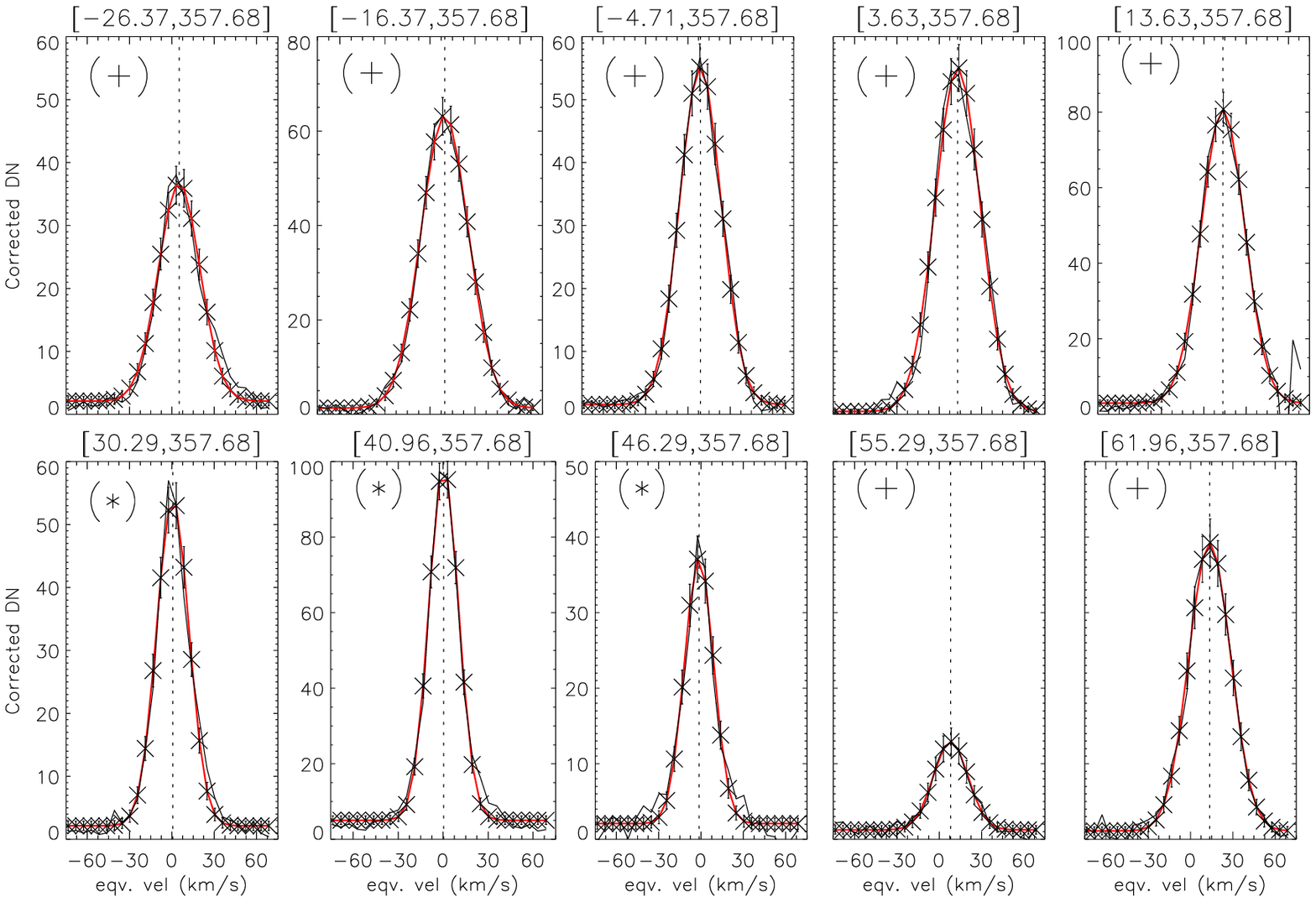}
\caption{Original spectral line profiles (in black) of \ion{Si}{4} 1394~{\AA} line in 3 $\times$ 3 pixel boxes around the locations marked with `+' and `*' in panel (C) of Fig.~\ref{march3}. The superimposed red curves show the corresponding Gaussian fits. The coordinates of these points in arcseconds are given at the top of each panel. The x-axis denotes the equivalent Doppler velocity scale. The error bars are also indicated.}\label{spectra}
\end{figure}

\begin{figure*}[!htbp]
\centering
\includegraphics[width= 1.0\textwidth]{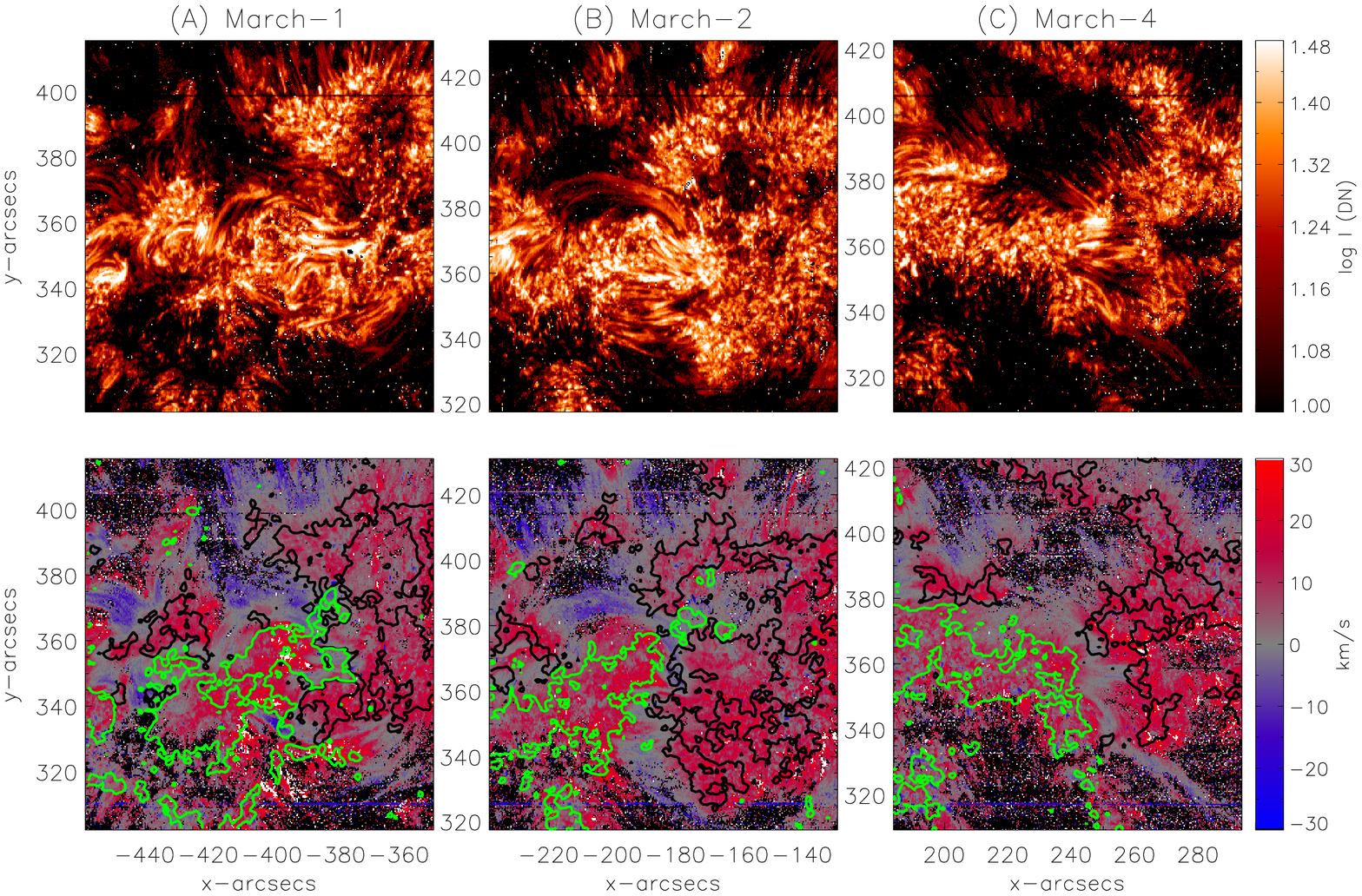}
\caption{Intensity (top row) and velocity maps (bottom row) for March 1, 2 and 4 obtained in \ion{Si}{4} 1394~{\AA} line. Over-plotted are the contours of LOS magnetic field between $\pm$50 G.}\label{velo1}
\end{figure*}

In order to understand how the line profiles vary across the active region, and in particular how they differ in strong and weak field regions, we have selected several locations along the horizontal stripe in Panel (C). The locations are chosen to sample the positive and negative strong field regions (identified by plus symbols) and the weak field corridor (identified by asterisks). Fig.~\ref{spectra} plots the spectral line profiles averaged within boxes of 3 $\times$ 3 pixels centred around these locations and converted to equivalent Doppler velocities. The spatial coordinates (in arcseconds) are noted at the top of each panel (panels marked with `+' and `*'). These plots show that there are no obvious differences in the overall shapes of profiles in strong and weak field regions, i.e., the profiles are all reasonably Gaussian, with no distinguishing asymmetries. Assuming photon shot noise, the errors on the profiles is limited to within $\pm$10{--}15\%.

\begin{figure*}
\centering
\includegraphics[width= 1.0\textwidth]{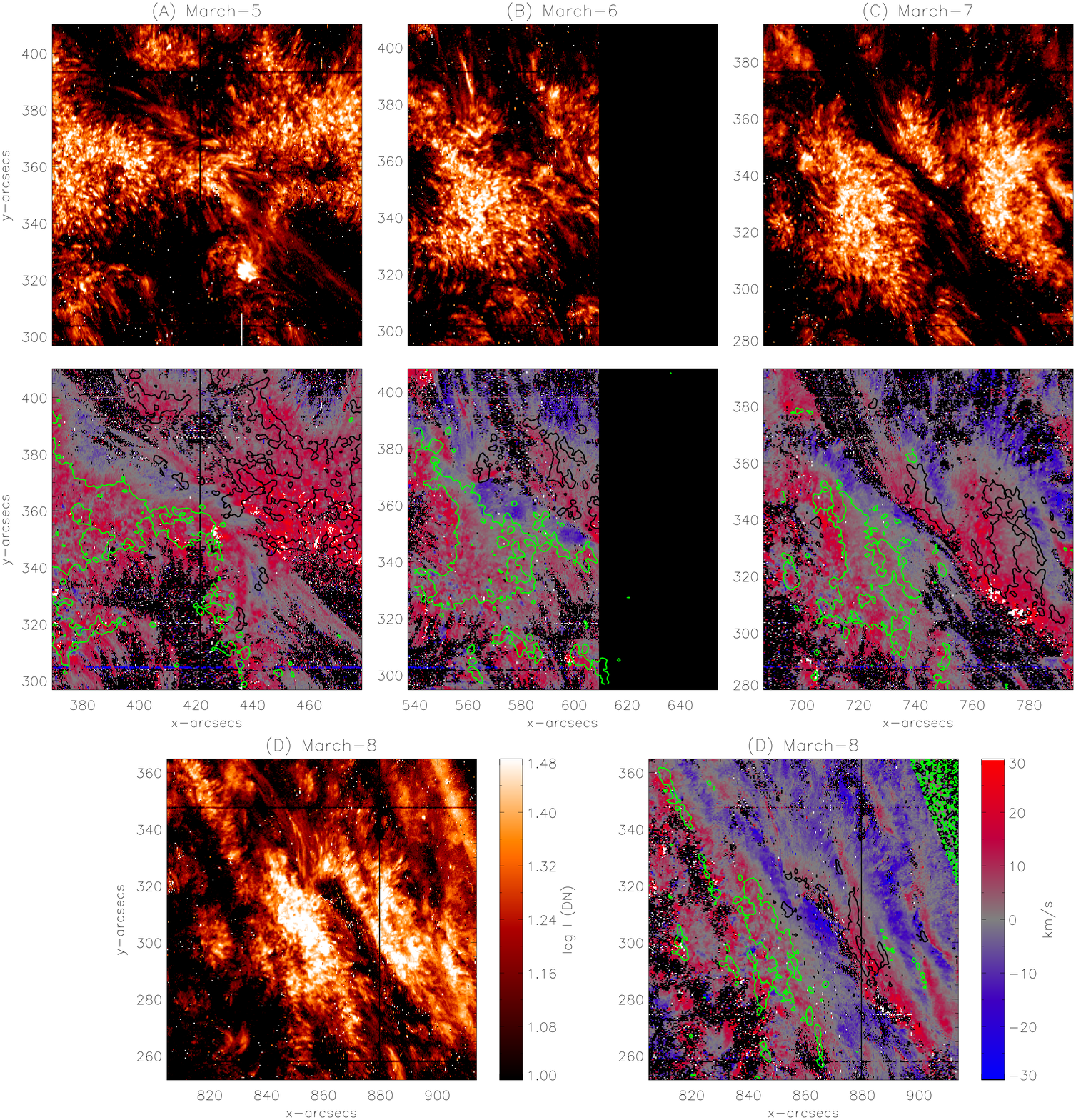}
\caption{Same as Fig.~\ref{velo1} for March 5, 6, 7 and 8.} \label{velo2}
\end{figure*}

\begin{figure}[t!]
\centering
\includegraphics[width= 1.0\textwidth]{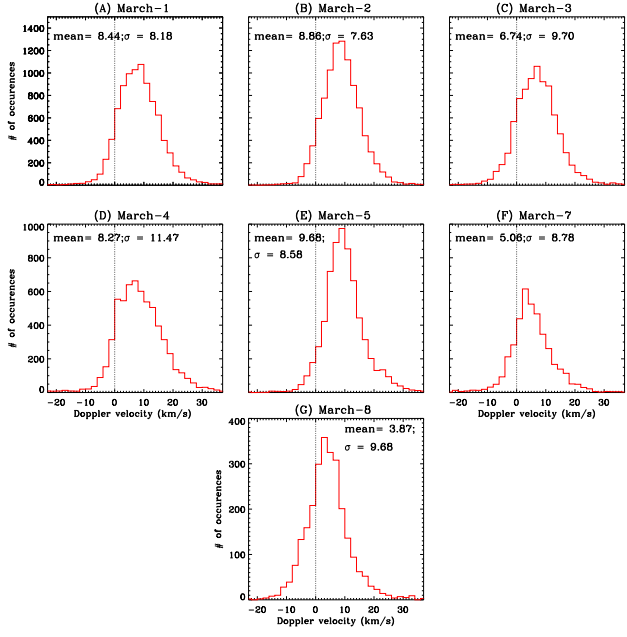}
\caption{Histograms of Doppler velocities in the strong field regions using \ion{Si}{4} line on the given dates. The mean and standard deviation values noted in each panel are in km~s$^{-1}$. The vertical dotted line marks the zero-velocity.} \label{hist_dv_ar}
\end{figure}

\begin{figure}[t!]
\centering
\includegraphics[trim=2.cm 0.cm 0.cm 0.cm,  width= 1.1\textwidth]{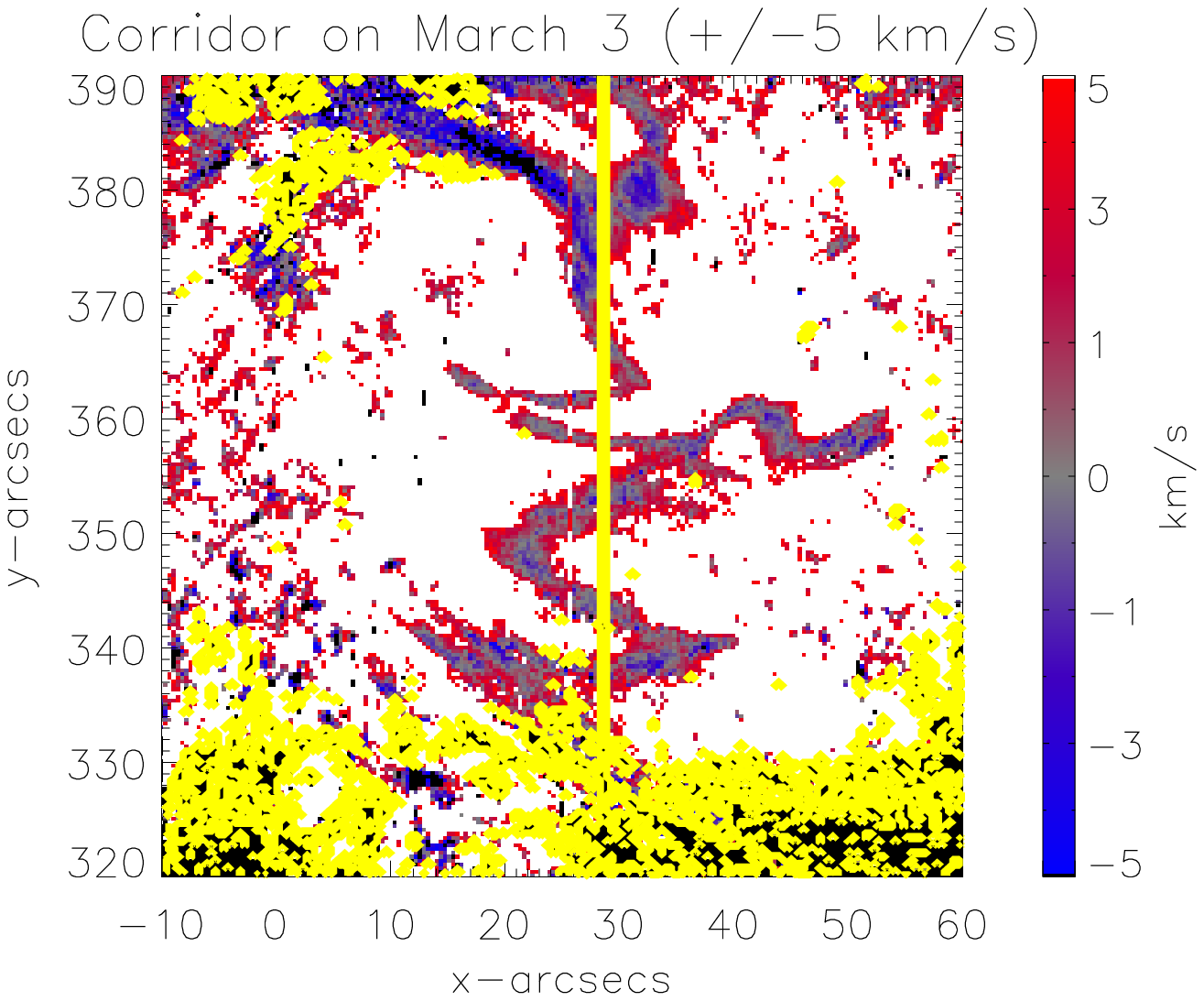}
\caption{A zoomed-in view of the corridor. All the missing pixels are shown in yellow whereas the white and black pixels show the saturated pixels in positive and negative velocities, respectively.}\label{corridor}
\end{figure}

Figs.~\ref{velo1} and \ref{velo2} display the intensity (top rows) and Doppler velocity maps (bottom rows) obtained for March 1, 2, 4, 5, 6, 7 and 8 over-plotted with the LOS magnetic field contours of $\pm$50~G. In the bottom panels the green contours represent positive polarity, whereas the black contours represent negative polarity. The Doppler velocity maps reveal that the strong field regions are predominantly redshifted throughout the entire duration of observation, as is also shown in the corresponding histograms plotted in Fig.~\ref{hist_dv_ar}. The obtained means of the distribution along with the cumulative random errors are given in Table~\ref{tab_dv}. In addition, there is a cumulative systematic error of 0.87~km~s$^{-1}$ (see Appendix~\ref{appen}). We note that in the velocity maps there are locations with strong field where the redshift reaches and even exceeds 20~km~s$^{-1}$. However the mean of velocities are only about 5{--}10~km~s$^{-1}$ (depending on the date of observation). We also note that the very small mean velocity obtained for March 6th is very likely due to the large chunk of missing exposure scans. In addition, there is a cumulative systematic error of 0.87~km~s$^{-1}$ on these estimates (for details, please see Appendix \ref{syst}).

\subsection{Doppler velocities in corridors}\label{dv_corr}

The weak field corridor separating the opposite polarity strong fields is more well-defined when the active region is closer to the disk center (corresponding to March 2/3). In order to have a detailed look within this region, in Fig.~\ref{corridor}, we have zoomed-in and plotted the Doppler velocity map obtained for March 3 with saturation at $\pm$5~km~s$^{-1}$. In the figure, missing pixels are represented by yellow. The blue and redshifted pixels with velocity amplitudes larger than 5~km~s$^{-1}$ are shown by black and white, respectively. We see a clear trend. There are very small blueshifts, or no shift at all, at the center of the corridor, with a transition to redshifts of increasing magnitude moving away from the center toward the strong field on either side. The blueshifts are consistent with zero given the uncertainty in absolute calibration.

\begin{figure}[hbtp]
\centering
\includegraphics[width= 1.0\textwidth]{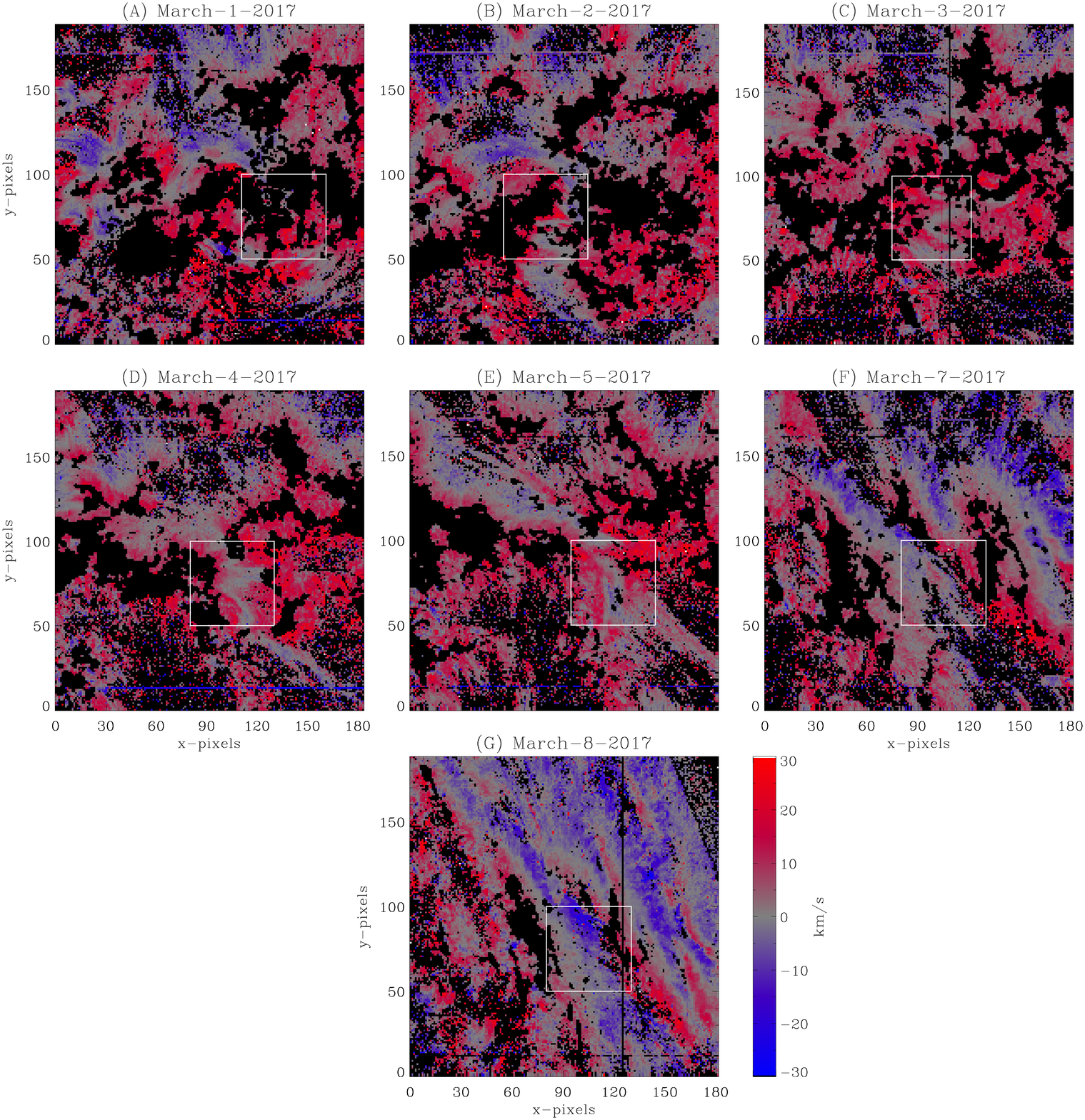}
\caption{Velocity maps over-plotted with boxes highlighting the corridor region. The strong field regions and saturated pixels are masked in black.} \label{corr_box}
\end{figure}

\begin{figure}[t!]
\centering
\includegraphics[width= 1.0\textwidth]{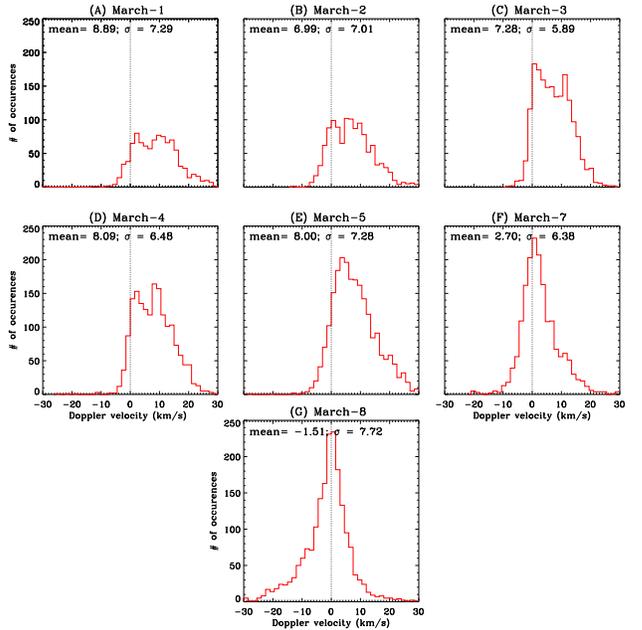}
\caption{Histograms of Doppler velocities in the corridor regions (shown within the white box in Fig.~\ref{corr_box}) using \ion{Si}{4} line on the given dates. The mean and standard deviation values noted in each panel are in km~s$^{-1}$. The vertical dotted line marks the zero-velocity.} \label{hist_dv_corr}
\end{figure}

\begin{figure*}
\centering
\includegraphics[width= 0.8\textwidth]{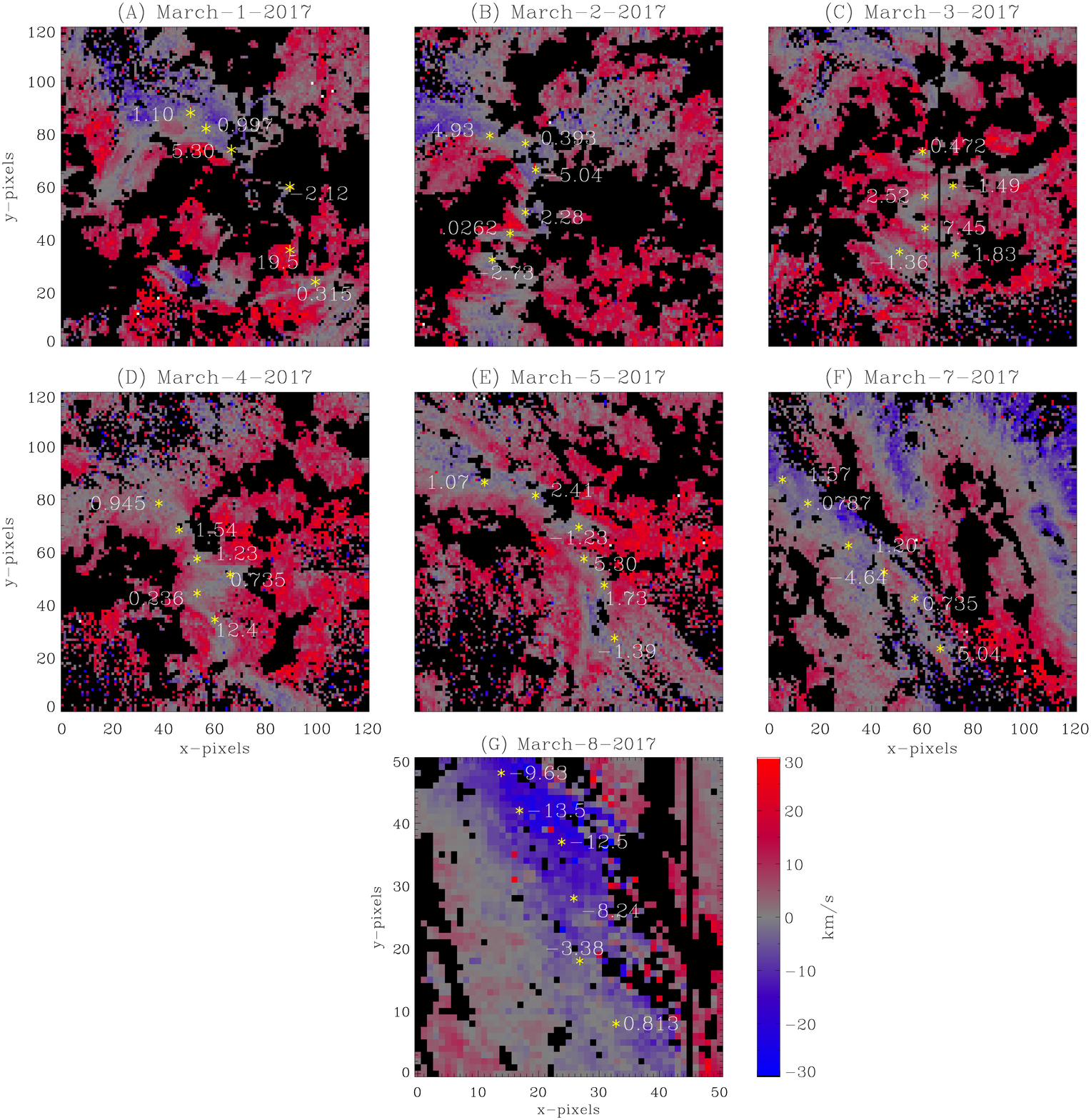}
\caption{Zoomed-in view of the corridor region (encompassing the white box in Fig.~\ref{corr_box}) to highlight the pixels having Doppler velocities very close to zero. A few such pixels are marked in yellow `*' with the corresponding velocities noted.} \label{profile_corr}
\end{figure*}

To measure the average velocities in the corridor, we limit ourselves to the small white box shown in Fig.~\ref{corr_box}. Pixels are included only if they fall within this box and have a field strength less than 50~G. In this figure, both strong field regions and saturated pixels (velocities larger than 30~km~s$^{-1}$) are masked in black. The distribution of velocities is shown separately for each day in Fig.~\ref{hist_dv_corr}. The obtained means along with the cumulative random errors of the distribution are given in Table~\ref{tab_dv}. In addition, there is a cumulative systematic error of 0.87~km~s$^{-1}$ on these estimates (see Appendix~\ref{appen}). It is interesting that the distributions appear to have two components, one centred very close to zero and another centred near 10~km~s$^{-1}$. The relative amplitudes of the two components seem to depend on disk position, with the 10~km~s$^{-1}$ component weakening closer to the limb. We find that, on average, the weak field corridor is redshifted by $\sim$3{--}9~km~s$^{-1}$ (excluding data on March 6 due the reason mentioned above). It is to be noted that the mean velocities in the corridor are similar to those in the strong field regions. It is tempting to suggest that the 10~km~s$^{-1}$ component is related to the lone component in the strong field distributions, with the corridor component near zero being physically separate. 

We note that some of the weak field pixels within the white boxes do not belong to the corridor proper. To better isolate the corridor velocities, we have identified particular pixels as shown with yellow asterisks in Fig.~\ref{profile_corr}. The measured velocities are indicated. This quantitatively verifies the point made earlier, that the center of the corridor has very small absolute velocities. It may be that there are two quasi-distinct classes of flow. One class is from  the central part of the corridor and is represented by the slow component of the corridor velocity distribution. The other class is from both the strong field regions and the outer part of the corridor and is represented by both the strong field velocity distribution and the fast component of the corridor velocity distribution (at around 10~km~s$^{-1}$).

\subsection{Center-to-Limb Variation of Doppler shifts}\label{dv_clv}

\begin{figure}[t!]
\centering
\includegraphics[trim=1.cm 4.cm 0.cm 6.cm,  width= 1.0\linewidth]{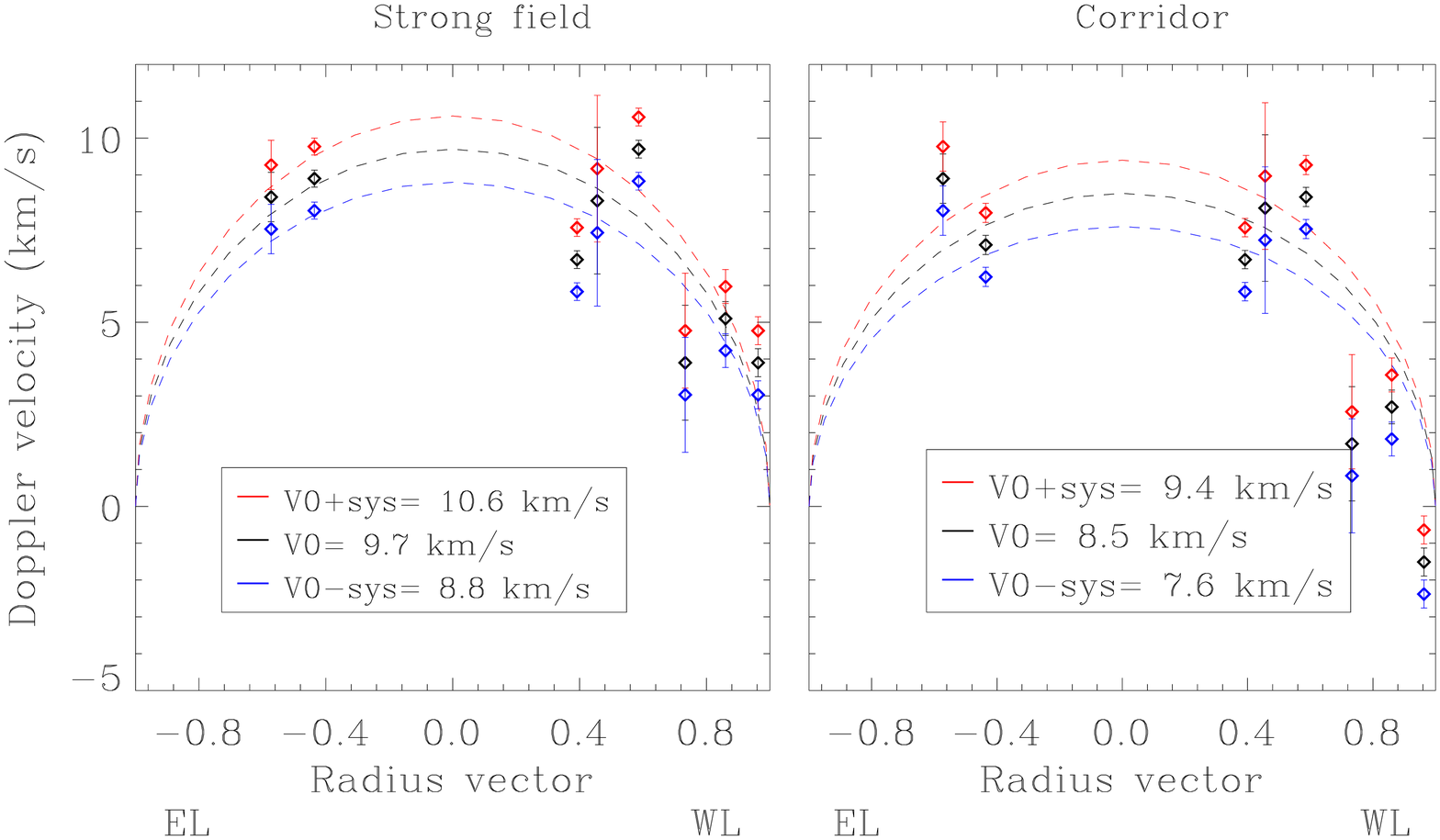}
\caption{Doppler velocities as a function of radius vector for strong field regions (left) and weak field corridors (right). The black diamonds represent the mean Doppler velocities without any systematic error (0.87~km~s$^{-1}$) taken into consideration. The corresponding uncertainties (random, in origin) ranging between $\pm$0.23{--}2.00~km~s$^{-1}$ are over-plotted. The red and blue diamonds represent the same but with the net systematic error added and subtracted, respectively. The dashed black lines correspond to the expected CLV (cosine function of heliographic longitude) for velocities 9.7 and 8.5~km~s$^{-1}$ measured at the disk center for the strong field regions and corridor, respectively. Correspondingly, the red and blue curves show the expected CLV curves for velocities with the systematic errors taken into account.}\label{theta}
\end{figure}

We have further studied the CLV of the Doppler velocities for strong field regions (left panel) and weak field corridors (right panel) in Fig.~\ref{theta}, where the observed mean LOS velocities are plotted (black diamonds) as a function of the radius vector. The radius vector is defined as the fractional distance of the feature to the limb from the disk center (following \cite{Kli_1987}). In this convention, a negative (positive) radius vector implies longitudes to the east (west) of the central meridian (represented by `EL' and `WL', respectively, in the plot). The vertical bars represent the cumulative random errors with the minimum and maximum being $\pm$0.23 and $\pm$2.00~km~s$^{-1}$, respectively. The red (blue) diamonds show the corresponding values of velocities when the total systematic error of 0.87~km~s$^{-1}$ is added (subtracted). The random errors on those estimates are also shown in red (blue) vertical bars. Over-plotted black dashed lines are the Doppler shifts expected for hypothetical vertical flows (V$_{los}= $ V$_{vertical}$ $\times$ $\sqrt{[1-(\frac{r}{R})^{2}]}$ where $\frac{r}{R}$ is the radius vector) with amplitudes of 9.7 and 8.5~km~s$^{-1}$ for strong field region and corridor, respectively. The red and blue dashed curves show the same with the systematic error taken into account, i.e. for 9.7$\pm$0.87 and 8.5$\pm$0.87~km~s$^{-1}$, respectively. We note that for the strong field regions we have a modest CLV. However, the interesting point arises in case of the corridors. Comparing this CLV plot for corridors with Figure~\ref{hist_dv_corr}, we note that the 10~km~s$^{-1}$ component in the histograms decreases significantly as we move closer to the limb. Therefore, the apparent large CLV in the average Doppler velocities in the corridors is due to the changing proportion of pixels in the two components, and not a real CLV. The actual CLV in the 10~km~s$^{-1}$ component is much weaker than the variation in the averages (as in seen for the pixels with strong magnetic field strengths). The following picture is suggested. There are two populations of flows which do not correspond precisely to the two populations of magnetic fields that we have defined (line-of-sight strength less than and greater than 50 G). The first population comes from the center of the weak field corridor and has Doppler velocities near zero. This is consistent with very small flows and horizontal magnetic fields. The second population comes from both the strong field regions and the outer parts of the weak field corridor. It has Doppler velocities near 10~km~s$^{-1}$ that decrease modestly from center to limb. Vertical downflows are suggested. However, the CLV is weaker than expected for a vertical flow. This is the dilemma pointed out by  \cite{Kli_1987}, as we discuss below.

\begin{figure}[t!]
\centering
\includegraphics[width= 1.0\textwidth]{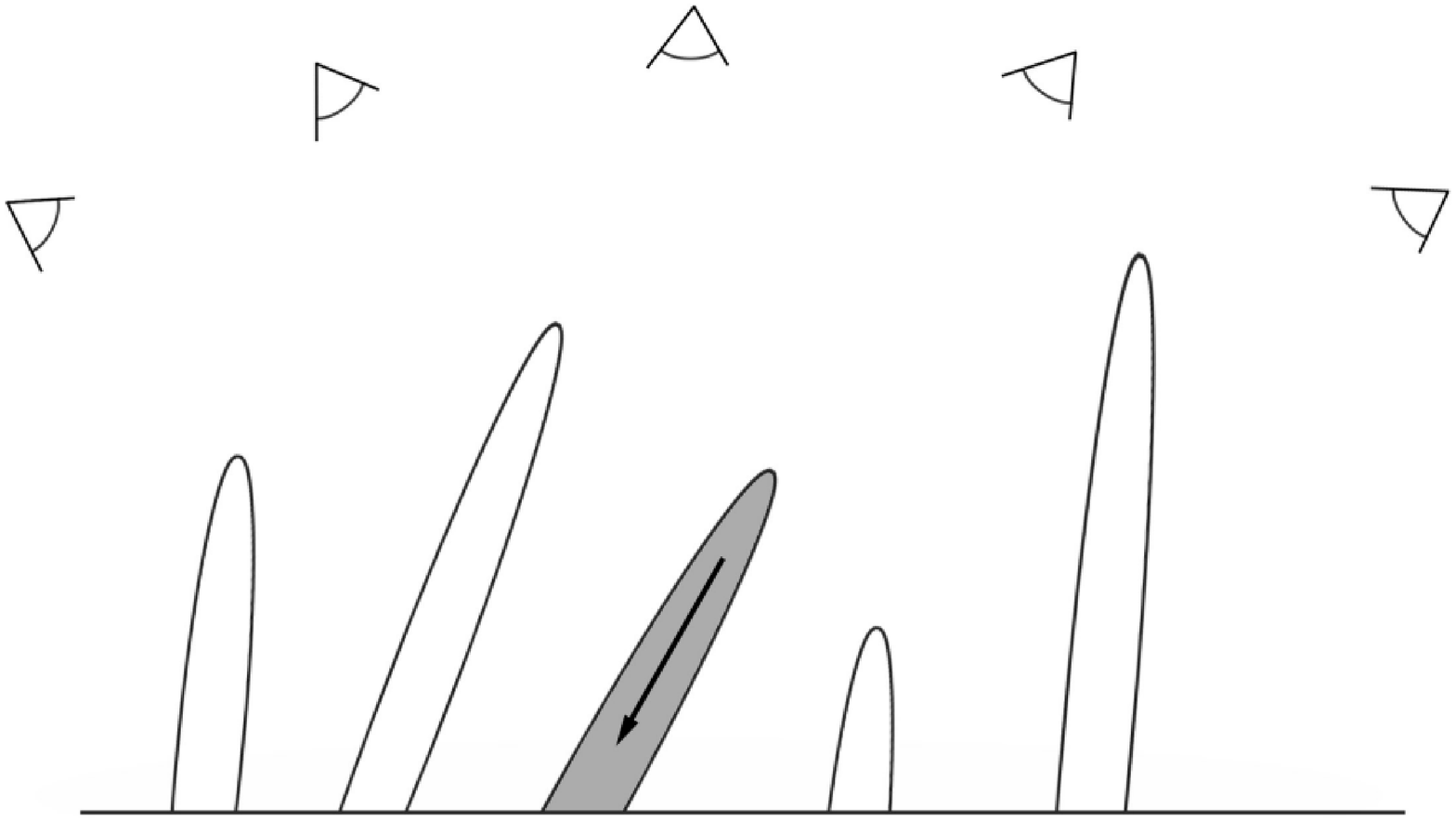}
\caption{A cartoon depicting the picture of visibility of spicules along the LOS.}\label{spicules}
\end{figure}

\section{Summary} \label{sum}

In this study we have measured the evolution of Doppler shifts in the lower transition region of an active region as it traversed across the central meridian. It is comprised of two strong, opposite polarity magnetic field regions and houses footpoints of loops which are visible in AIA 171~{\AA} channel images. There is an intermediate weak field corridor. The strong (weak) polarity regions were defined as regions with magnetic field stronger (weaker) than 50~G. The corridor region evolved and grew wider as the active region evolved further. For the study, we have used \ion{Si}{4} ($\log\,T[K] =$ 4.9) observed with IRIS. Moreover, we have compared the structure of Doppler shift with the photospheric magnetic field using the observations taken from HMI. Below we summarise our findings.

\begin{enumerate}

\item We find that the \ion{Fe}{2} 1392.8~{\AA} is not suitable for absolute velocity calibration because of its blue wing asymmetry, particularly near the limb that could be related to opacity effects \citep{FelDP_1976, DoyW_1980, ErdDP_1998}. Since \ion{O}{1} 1355.6~{\AA} has no such variation or asymmetries, we have used it for deriving the absolute Doppler velocities. 

\item Our analyses reveal that there is no conspicuous location dependent difference in the shape of the spectral line profile of \ion{Si}{4} in the form of asymmetry or any excess broadening irrespective of the magnetic field strength in the region.

\item The Doppler maps obtained using \ion{Si}{4} at $\log\,T[K] =$ 4.9 demonstrate that strong field regions are predominantly redshifted with average velocities of $\sim$5{--}10~km~s$^{-1}$. However, in the maps there are regions with stronger velocities such as $\sim$20{--}30~km~s$^{-1}$.

It is pertinent to compare these with velocities in moss regions which are observed in EUV lines, showing redshifts of the order of 5~km~s$^{-1}$ in \ion{Fe}{8} \citep{TriMK_2012} and \ion{Ni}{8} \citep{WinTM_2013}.

\item The average velocities in the weak field corridor range between $\sim$3{--}9~km~s$^{-1}$. A detailed look into the weak field corridor suggests that there is a narrow lane of near-zero velocity at the center, along with redshifted pixels near the boundary between the corridor and strong field regions. In addition, there are isolated patches of blueshifts.

\item Within the moderate range of longitudes studied here and an error margin of $\pm$0.23{--}2.00~km~s$^{-1}$ in both regions, there is a moderate evidence for CLV of Doppler velocities in strong field regions. In the weak field corridors, we note that there are two velocity components- a zero-velocity component and one at around 10~km~s$^{-1}$. The former comes from the pixels very close to the center of the weak field corridors and hence, have very small flows along the LOS. The latter is noted to have a modest CLV, similar to that for the case of the strong field regions. However, this CLV is weaker than that expected for vertical flows. This is true irrespective of whether the systematic errors are added or subtracted or ignored from the estimates.

\end{enumerate}

\section{Discussions} \label{disc} 

Using data from UVSP on SMM, \cite{Kli_1987, Kli_1989} showed that \ion{C}{4} (1548~{\AA})  is relatively redshifted in strong field regions and relatively blueshifted in weak field corridors, irrespective of position on the disk. Since there was no absolute wavelength calibration, the Doppler shifts were measured relative to the average Doppler shift over the entire 4$\arcmin \times$4$\arcmin$ FOV. Klimchuk pointed out that the magnetic field in the upper photosphere, chromosphere, and transition region of corridors is largely horizontal as a consequence of the well-known canopy effect, first emphasized by \cite{Gio_1980}, in which concentrated fields expand dramatically with height as they emerge from the high-$\beta$ environment of the photosphere into the low-$\beta$ environment of the corona. Since plasma is constrained to flow along the magnetic field in the transition region, any \ion{C}{4} flows at the centers of corridors are expected to be horizontal. This led Klimchuk to suggest that the relative blueshifts observed by UVSP near disk center, where the LOS is vertical, actually correspond to absolute Doppler shifts near zero. However, these same horizontal flows would produce sizable Doppler shifts when observed near the limb. A blueshift (redshift) near the east limb should weaken to zero as the active region rotates to central meridian, and then become a redshift (blueshift) of increasing amplitude as the region continues to rotate toward the west limb. Klimchuk therefore went on to suggest that the relative blueshifts in corridors likely correspond to absolute Doppler shifts near zero at {\it all} disk positions, implying flows that are very slow. Since \ion{Si}{4} is formed at a similar temperature to \ion{C}{4}, our measurements of small absolute Doppler shifts in the center of the corridor support this suggestion.

Klimchuk noted a puzzling consequence of his proposed absolute Doppler shift calibration. When the relative blueshifts in the corridor were set to zero, the redshifts in the strong field regions ended up having a similar amplitude all across the disk. This is challenging to understand. When redshifts are seen in both the eastern and western hemispheres, the most obvious explanation is that the flows are approximately vertical (downward). However, the redshifts produced by vertical downflows should decrease in amplitude from center to limb, as indicated by the dashed curves in Fig.~\ref{theta}. The absolute redshifts we measure in the strong field regions and high-velocity component of the corridor have greater CLV than inferred by Klimchuk, but still less than expected for a vertical flow. This is also true of the measurements by \cite{FelDC_1982}.


How are we to solve this dilemma? \cite{Ant_1984} proposed a clever explanation. He introduced the concept of a chromospheric `well' to explain the constancy of redshifts across the disk. The idea is that downflows occur in the legs of some magnetic strands that have enhanced pressure relative to their surroundings. The chromosphere of these strands will be locally depressed, causing the transition region to sit at the bottom of what is essentially a well. If strands have a range of inclinations relative to vertical, then the transition region downflows will be visible only in those strands where the LOS is approximately aligned with the strand axis, i.e., when looking `down' the well. If all downflowing strands have a similar velocity distribution, independent of their inclination, then the observed redshifts will be similar everywhere on the disk.

Impulsive heating is the mostly likely cause of the downflows. It produces both upflows during an evaporation phase and downflows during a draining phase; however, the upflows are shorter lived and fainter, and they contribute far less to the time averaged emission (or spatially averaged emission for an unresolved collection of strands) \citep{PatK_2006}. Another advantage of impulsive heating is that it provides the enhanced pressures necessary to depress the chromosphere and transition region into the well.

As appealing as this picture is, there is a significant discrepancy with observations: the redshifts predicted for the lower transition region with impulsive heating are much slower than observed. Hydrodynamic (loop) models give redshifts of 1{--}3~km~s$^{-1}$ in \ion{Fe}{8} (185~{\AA}), which is formed at approximately 0.4~MK \citep{TarB_2014, LopK_2018}. For constant mass flux at constant pressure, velocity is proportional to temperature, so the corresponding redshifts in \ion{Si}{4} and \ion{C}{4} should be $<$ 1~km~s$^{-1}$. This is an order of magnitude slower than observed. 

It has long been known that emission cooler than about 0.1~MK is far brighter than predicted by standard loop models, whether the heating is steady or impulsive. \cite{AntN_1986} suggested that the emission comes not primarily from the lower transition region of hot loops, but rather from small, low-lying cool loops. Such loops can exist in regions of highly mixed magnetic polarity, so this is a plausible explanation for the quiet Sun. However, small loops are not present in the large unipolar areas of active regions, and an alternative explanation is necessary.

It was discovered not long ago that a subset of spicules are heated to $\sim$0.1~MK as they rise, with the tips being heated to even higher temperatures \citep{DePMC_2011}. This led \cite{Kli_2012} to suggest that these `type II' spicules are the primary source of \ion{C}{4}/\ion{Si}{4} emission in active regions and possibly also the rest of the Sun. Most of the type II spicule mass falls back to the surface after being ejected \citep[also see,][]{PneK_1978, AthH_1982, Ath_1984}. Only the hot tip continues to rise into the corona. Since 0.1~MK emission is produced during the entire falling phase, but only during a fraction of the rising phase, a net redshift would be expected.


Alternatively, we suggest an idealized concept similar to the chomospheric well. As with the well, obscuring plasma limits the visibility to type II spicules that are approximately aligned with the LOS. Instead of a well, however, we propose a `wall' of cold ($\sim$10$^{4}$~K) spicules extending upward from the chromopshere. Most spicules are of this type. They are sometimes referred to as classical spicules and sometimes as type I, although spicule classification is an ongoing debate \citep{RaoPP_2016}. Fig.~\ref{spicules} is a cartoon of the picture we have in mind showing the visibility of the spicules along the LOS at several disk positions.

To produce a wall that is effective at obscuring a type II spicule, the cold type I spicules must have a significant optical thickness at 1394~{\AA}, the wavelength of \ion{Si}{4}. Continuum opacity from the ionization of ground state hydrogen and helium is known to be strong, but it occurs at shorter wavelengths and plays no role here. There are, however, other sources of significant opacity at 1394~{\AA} (ionization from other species and from excited states of H and He).

\cite{Rut_2016} has computed the extinction of \ion{Si}{4} (1394~{\AA}) from cold, dense solar plasmas in thermodynamic equilibrium (TE). His Figure 5 covers temperatures from 3000 to 43,000~K and hydrogen densities from 10$^{13}$ to 10$^{16}$~cm$^{-3}$. He has kindly provided a corresponding plot for n$_{H}$ = 10$^{11}$~cm$^{-3}$ (R. Rutten, private communication). \cite{Bec_1972} gives the following base-to-tip physical conditions of a typical type I spicule:  T $=$ 9000{--}16,000 K, n$_{H}$ = 1.6$\times$10$^{11}${--}3.4$\times$10$^{10}$~cm$^{-3}$ and diameter $=$ 10$^{8}$~cm. We find that such a spicule, if in thermodynamic equilibrium, would have an optical thickness approaching and even exceeding unity. In reality, the ionization and excitation state of the plasma will be intermediate between that given by the Saha and Boltzmann equations, as applies in TE, and so-called `coronal equilibrium' where there is a balance between collisional excitation and radiative de-excitation. The true opacity of type I spicules at 1394~{\AA} will therefore be less than what we have estimated. It may nonetheless be enough to explain the reduction in CLV relative to vertical flow. Some variation is seen, so the type I spicules need not be opaque.

\begin{appendix}
\section{Appendix: Error estimates on Doppler shift} \label{appen}

The average velocities derived using IRIS incurs both random as well as systematic errors. Below we describe these errors incurred in this study in detail.

\subsection{Random errors}\label{ran}
\begin{itemize}

\item One of the first random error is standard error on velocities, which is defined as the $E_{1} = \frac{\sigma}{\sqrt N}$, where $\sigma$ is the standard deviation of the velocity errors at the individual pixels and N is the total number of pixels. 

\item The second random error is related to the difference between in the central wavelength derived from the fitted spectral curve of raster-integrated \ion{O}{1} line, which we assign to the adopted rest wavelength of 1355.598~{\AA}, and the true position of this wavelength. The error, which we call E2, arises from photon shot noise, inaccuracies in the fitting, and real deviations of the average O I velocity from true zero.



\end{itemize}

Both of these vary from one IRIS raster to another. Since these two are independent errors, the cumulative random error (S$_{ran}$) is obtained as \[\mathit S_{ran} = \sqrt {(E_{1}^2 +  E_{2}^2)     }\]

\subsection{Systematic errors}\label{syst}

\begin{itemize}
\item A systematic error of 3~m{\AA} (0.66~km~s$^{-1}$) is incurred in all Doppler velocity measurements because of uncertainty in the true rest wavelength of the \ion{O}{1} calibration line. An estimate of this uncertainty comes from differences between the line position determined from HRTS observations around the limb and the laboratory wavelength of 1355.598~{\AA} \citep{SanB_1986}. We call this as E$_{3}$.

\item The second systematic error is introduced due to uncertainty in dispersion which is of about 0.1 pixel. Given that the spectral resolution of FUV IRIS is 26~m{\AA}/pixel, the resultant uncertainty is \\ (0.1*26*10$^{-3}$/1393.755)*3*10$^{5}$ =  0.56~km~s$^{-1}$ for the \ion{Si}{4} line at 1393.755~{\AA}. We call this as E$_{4}$.

\end{itemize}
 
The above two systematic errors are independent, the cumulative systematic error (S$_{sys}$) is obtained as \[\mathit S_{sys} = \sqrt {(E_{3}^2 +  E_{4}^2)     }\]

\end{appendix}

\begin{acknowledgements}

This research is supported by the Max-Planck India Partner Group of MPS at IUCAA. AIA and HMI data are courtesy of SDO (NASA). IRIS is a NASA small explorer mission developed and operated by LMSAL with mission operations executed at NASA Ames Research center and major contributions to downlink communications funded by ESA and the Norwegian Space Centre. The work of J.A.K. was supported by the NASA Supporting Research program and GSFC Internal Scientist Funding Model (competitive work package) program. We thank Robert Rutten, Han Uitenbroek and Adrian Daw for helpful discussions concerning spicule opacity. We also thank Adrian Daw and Don Schmit for their help in understanding the thermal orbital variations of IRIS.

\end{acknowledgements}

\bibliographystyle{apj}
\bibliography{references}

\begin{center}
\vspace{1cm}
\captionof{table}{Detailed information regarding the dates of observations, the reference wavelengths derived on each date based on the centroid of the Guassian fitting of the \ion{O}{1} line profiles and the mean Doppler velocities measured with \ion{Si}{4} 1393.7~{\AA} line. These estimates are done separately for strong field regions and corridor (highlighted within the white box in Fig.~\ref{corr_box}). The corresponding cumulative random errors are also quoted. The rasters on March 6 have a lot of missing exposures and that on March 7{--}8 are very close to the limb, hence the results could not be realistic estimates.}\label{tab_dv} 
\label{details}
\vspace{1cm}
\begin{tabular}{| c | c | c | c |}  \hline

Date     &$\lambda$  &Strong field   & Corridor  \\
 2017       &\ion{O}{1} ({\AA}) &(km s$^{-1}$)   & (km s$^{-1}$) \\ \hline
& &Mean    &Mean   \\  \hline
1 Mar   &1355.601 &8.4$\pm$0.67 &8.9$\pm$0.67 \\
2 Mar &1355.599  &8.9$\pm$0.23 &7.0$\pm$0.26 \\
3 Mar  &1355.599 &6.7$\pm$0.24 &7.3$\pm$0.25  \\
4 Mar  &1355.607 &8.3$\pm$1.99 &8.1$\pm$1.99  \\
5 Mar  &1355.599 &9.7$\pm$0.24 &8.0$\pm$0.26 \\
6 Mar  &1355.605 &3.9$\pm$1.56  &1.9$\pm$1.55 \\
7 Mar  &1355.600 &5.1$\pm$0.46  &2.7$\pm$0.46 \\
8 Mar  &1355.599 &3.9$\pm$0.38  &-1.5$\pm$0.38 \\ \hline
\end{tabular}
\end{center}

\end{document}